\documentclass{article}
\usepackage{spconf,amsmath,graphicx,multirow}
\usepackage[dvipsnames]{xcolor}
\usepackage{makecell}
\usepackage{amsmath}
\usepackage{arydshln}
\usepackage{algorithm}
\usepackage{algpseudocode}

\title{Exploring the Viability of Synthetic Audio Data for Audio-Based Dialogue State Tracking}

\name{Jihyun Lee$^{*,1}$, Yejin Jeon$^{*,1}$, Wonjun Lee$^{2}$,  Yunsu Kim$^{3}$, Gary Geunbae Lee$^{1,2}$}
\address{$^1$Graduate School of Artificial Intelligence, POSTECH, Republic of Korea\\
  $^2$Department of Computer Science and Engineering, POSTECH, Republic of Korea \\
  $^3$ aiXplain Inc. Los Gatos, CA, USA\\
  \{jihyunlee, jeonyj0612, lee1jun, gblee\}@postech.ac.kr, yunsu.kim@aixplain.com}

\copyrightnotice{979-8-3503-0689-7/23/\$31.00~\copyright2023 IEEE}
\begin{document}
%
\maketitle
\begin{abstract}

Dialogue state tracking plays a crucial role in extracting information in task-oriented dialogue systems. However, preceding research are limited to textual modalities, primarily due to the shortage of authentic human audio datasets. We address this by investigating synthetic audio data for audio-based DST. To this end, we develop cascading and end-to-end models, train them with our synthetic audio dataset, and test them on actual human speech data. To facilitate evaluation tailored to audio modalities, we introduce a novel PhonemeF1 to capture pronunciation similarity. Experimental results showed that models trained solely on synthetic datasets can generalize their performance to human voice data. By eliminating the dependency on human speech data collection, these insights pave the way for significant practical advancements in audio-based DST. Data and code are available at https://github.com/JihyunLee1/E2E-DST. \footnote{*The first two authors contributed equally to this work.}

\end{abstract}
\section{Introduction}
\label{sec:intro}

\begin{figure}[t]
  \centering
  \includegraphics[width=65mm]{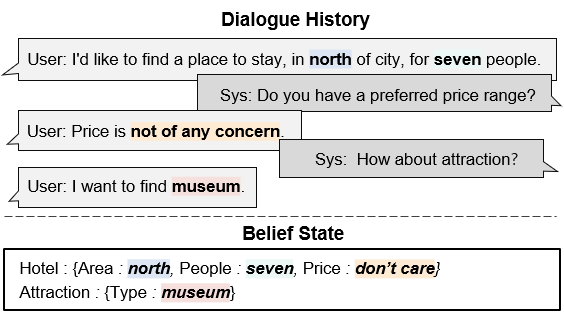}
  \caption{An example dialogue in the DST dataset. Given the dialogue history, the DST model predicts a belief state, which consists of slots (e.g., Hotel-Area) and its values (e.g., north).}
  \label{fig:DST}
\end{figure}

Dialogue state tracking (DST) models play an essential role in facilitating task-oriented communication by extracting useful information from user-system conversations \cite{young2013pomdp}. The extracted information is called the \textit{belief state}, and it consists of a slot  and its value pair (Figure~\ref{fig:DST}). Traditionally, DST has primarily focused on textual data \cite{trade,chao2019bert,gao2020machine}, and hence does not consider other modalities such as speech. However, with the increasing prevalence of voice-based conversational interfaces, there is a growing demand to extend text-based DST to the audio domain.


Audio-based DST offers several advantages in enhancing user experience. First, it enables more natural and intuitive communication through voice commands, broadening its appeal to a wider audience \cite{chen2018construction}, and facilitates multitasking scenarios, such as driving or cooking \cite{parush2005speech}. In addition, audio-based DST can leverage audio-specific features and cues, such as intonation and speaker characteristics, which aid in understanding intended meaning \cite{prosody}.

Nonetheless, a significant challenge in developing audio-based DST systems lies in the availability and collection of real human voice data. Gathering large-scale and diverse voice datasets can be costly, time-consuming, and may involve privacy concerns. In addition, annotating belief state labels in audio data prove labor-intensive and demanding. To address these limitations, exploring synthetic audio data as an alternative approach holds substantial promise.

Synthetic audio data offers an attractive solution for training audio-input DST models without relying on actual human voice data. By leveraging text-to-speech (TTS) models \cite{Intro-1, FastSpeech2}, it becomes possible to generate a diverse and customizable synthetic audio dialogue dataset that covers a wide range of scenarios and contexts. Such synthetic audio datasets can be created easily and in a cost-effective manner, which eliminates the need for extensive human voice data collection.

This paper addresses the research question of whether synthetic audio data can serve as a viable alternative to authentic human data while yielding promising results in audio-based DST. Towards this end, we generate a synthetic audio dialogue dataset (SynthWOZ) by combining existing textual dialogue datasets (i.e., MultiWOZ2.1 \cite{MWOZ}) with TTS. We carefully preprocess the transcripts to ensure optimal audio quality and clarity, and curate a multi-speaker synthetic dataset for better generalizability to actual human voices. Moreover, we develop two baseline models for speech-based DST: a cascading model that incorporates an Automatic Speech Recognition (ASR) module that is followed by a text-based DST module, and an end-to-end (E2E) model that directly maps input audio dialogues to belief states.

Furthermore, we introduce a novel evaluation metric called PhonemeF1, which is designed specifically to capture pronunciation similarity in the context of DST in the audio domain. Previous DST research has primarily focused on text-based evaluation metrics, which overlooks the unique characteristics of spoken utterances. Our PhonemeF1 metric bridges this gap by giving partial credit to predictions with higher pronunciation similarity with that of the ground truth, allowing for more reasonable scoring.


Through experiments, we were able to observe promising results when models that are trained solely on synthetic speech, are adapted to actual human speech. Furthermore, we also conduct analyses of the error patterns that arise for both our cascading and E2E models. As our research demonstrates the feasibility of utilizing cost-effective and readily available synthetic audio data, we hope to provide a foundation for the development of effective and practical audio-based DST systems using synthetic corpora.


\section{Related Work}
\noindent
\noindent
\newline
\textbf{Leveraging Synthetic Data} TTS-synthesized datasets has emerged as a viable method in low-resource environments and various research domains. Through augmentation of authentic human datasets using TTS, not only does this enhance the accuracy of self-training processes \cite{Related-self}, but also the generalizability of ASR models by introducing prosodic and acoustic variations \cite{Related-AugmentTTS, Related-AugmentTTS2} into training data. Furthermore, as recent TTS models \cite{Intro-1, FastSpeech2} are able to produce more natural, human-like speech, this has prompted researchers to explore synthetic speech datasets as complete substitutes for authentic human speech datasets \cite{Related-punc, Related-JustTTS}. Nevertheless, despite these notable developments in other research areas, no prior investigations have explored the potential impact of using synthetic data in DST. Therefore, our study provides a comprehensive investigation of the influence of synthetic datasets for audio-based DST.
\newline
\newline
\noindent
\textbf{Comparisons Between Related Tasks}
Significant progress has been made in developing automated systems to support user intent understanding. For instance, task-oriented dialogue research has received considerable attention usually within the text modality \cite{MWOZ}, leading to the emergence of sub-areas such as dialogue state tracking (DST) \cite{trade,chao2019bert,gao2020machine}, policy determination \cite{pol-1, pol-2, pol-3}, and dialogue generation \cite{gen-1}. Within the audio modality, spoken language understanding (SLU) models primarily focus on single-turn user commands by extracting relevant information to fulfill the intended actions \cite{Related-SLU-1, Related-SLU-2}. Intent classification is another prominent research area that has been extensively explored across text \cite{intent-text}, speech \cite{intent-audio}, and multi-modal environments \cite{intent-multi}. While audio-based DST shares some similarities with these aforementioned tasks, it also exhibits notable distinctions; it addresses acoustic inputs that are comprised of multi-turn interactions between users and systems, and requires the prediction of comprehensive information that is beyond simple categorization of user intent.


\section{Preliminaries}



This study focuses on the accurate extraction of slot-value pairs from a multi-turn conversation, which is specifically conducted in the audio modality. Let conversation history $C$ at time step $t$ be denoted as $C_t = \{u_1,s_1,...,s_{t-1},u_t\} $, where $u_{t}$ represents the user's speech and $s_{t}$ represents the response from the system. The turn-level belief state $b_{t}$ at turn ${t}$ comprises slot-value pairs. The sequence of belief states at turn $t$, denoted as $B_t = \{b_1, b_2, ..., b_t\}$, represents the accumulation of the beliefs up to turn $t$, where each $b_i$ corresponds to a turn-level belief state at turn $i$.

\section{SynthWOZ Dataset}
\subsection{Transcript Generation and Pre-Processing}
SynthWOZ is a comprehensive multi-turn synthetic audio dataset. Its creation involved utilizing the text transcripts required for audio generation, which were obtained from MultiWOZ 2.1 \cite{MWOZ}, a widely acknowledged benchmark resource for text-based DST. The dataset encompasses dialogues from seven different domains\footnote{Hotel, restaurant, taxi, train, attraction, hospital, and police.}. To improve coherence of our synthetic dataset, we perform preprocessing with the following procedures: (1) normalization of special characters (e.g., \$5$\rightarrow$five dollars), (2) substitution of numerical values with their corresponding alphabetic representations (e.g., 23$\rightarrow$twenty three), (3) unification of randomized sequences of identification from various ontologies into a single representation (e.g., postcode cb17ag $\rightarrow$ [number]), (4) rectification of misspellings, and (5) correction of erroneous word orderings (e.g., ``I the am" $\rightarrow$ ``I am the").

\subsection{Audio Dataset Generation}


Synthesis of audio samples for our SynthWOZ dataset was accomplished by leveraging the FastSpeech2 \cite{FastSpeech2} TTS and HifiGAN \cite{HifiGan} vocoder. These systems were trained on the widely recognized LibriTTS \cite{LibriTTS} corpus, while the target text inputs were derived from the preprocessed transcript described in Section 4.1. To enhance the dataset's resemblance to authentic human speech and promote its generalizability, we employed multiple speaker voices during the synthesis process by passing speaker labels to a fixed lookup table. Specifically, ten distinct voices, distributed equally across genders, were utilized. Furthermore, as audio-based DST is conducted in a multi-turn spoken dialogue setting, a crucial aspect involves understanding the previous system utterance. To illustrate, consider a user statement like \textit{``Ok, I will choose the first option."} Deciphering the meaning of the ``first option" requires access to the dialogue history, including the preceding system utterance. For this purpose, we ensured the synthesis of both system and user utterances, and employed a separate female voice to represent the system.




\begin{table}[t]
\begin{center}

\scalebox{0.80}{
\begin{tabular}{c|c|c|c|c|c}
\hline
\multirow{2}{*}{\textbf{Datasets}} &  \multirow{2}{*}{\textbf{Size}} &\multicolumn{2}{c|}{\textbf{W2V}}  & \multicolumn{2}{c}{\textbf{Whisper}} \\ \cline{3-6}
& & \multicolumn{1}{c|}{WER}  & CER  & \multicolumn{1}{c|}{WER}   & CER  \\ \hline
\textbf{Train}  & 8000     & \multicolumn{1}{c|}{9.57}  & 3.66 & \multicolumn{1}{c|}{8.58} & 3.78  \\
\textbf{Dev}    & 1000    & \multicolumn{1}{c|}{7.73}  & 2.01& \multicolumn{1}{c|}{7.13} & 2.18  \\
\textbf{Test}   & 120     & \multicolumn{1}{c|}{7.72}  & 3.85 & \multicolumn{1}{c|}{6.87} & 3.92  \\
\textbf{Test-Human} & 120  & \multicolumn{1}{c|}{15.06} & 6.23 & \multicolumn{1}{c|}{8.51} & 3.43 \\ \hline
\end{tabular}}
\caption{SynthWOZ data sizes and WER, CER scores. } 
\label{tab:human_test_ASR}
\end{center}

\end{table}

\subsection{Human Speech Dataset}
In addition to synthetic audio data, our SynthWOZ dataset includes authentic recordings of human speech. The inclusion of authentic human recordings allows us to assess the performance of models trained solely on synthetic data in practical scenarios that involve real human speech. Towards this end, we recruited a total of ten speakers (six females and four males) who come from diverse demographic backgrounds; four speakers have American/English accents, one is Germanic, and five have East and South Asian accents. These participants were instructed to engage in natural dialogue, acting as users interacting with the system\footnote{For ethical purposes, we informed the speakers of the intended use of the recordings prior to recording and obtained written consent for their open-sourced distribution for research purposes only.}. Moreover, the recordings were conducted in various environmental settings (i.e., microphones, headphones, and phones) in order to introduce environment noise into human audio, and closely replicate real-life scenarios.

\subsection{Audio Quality Assessment }


To test the quality of our SynthWOZ and actual human speech, we adopted the pre-trained Wav2Vec 2.0 (W2V) \cite{w2v} and Whisper \cite{whisper} models, which are state-of-the-art ASR systems trained on English and multilingual data, respectively. The accuracy of the resulting ASR transcriptions are calculated via Word Error Rate (WER) and Character Error Rate (CER). Results in Table~\ref{tab:human_test_ASR} demonstrate that there are minor pronunciation errors for both synthetic (i.e., \textit{Train, Dev, and Test}) and human speech (i.e., \textit{Test-Human}) datasets.

\begin{figure}[t]
  \centering
  \includegraphics[width=65mm]{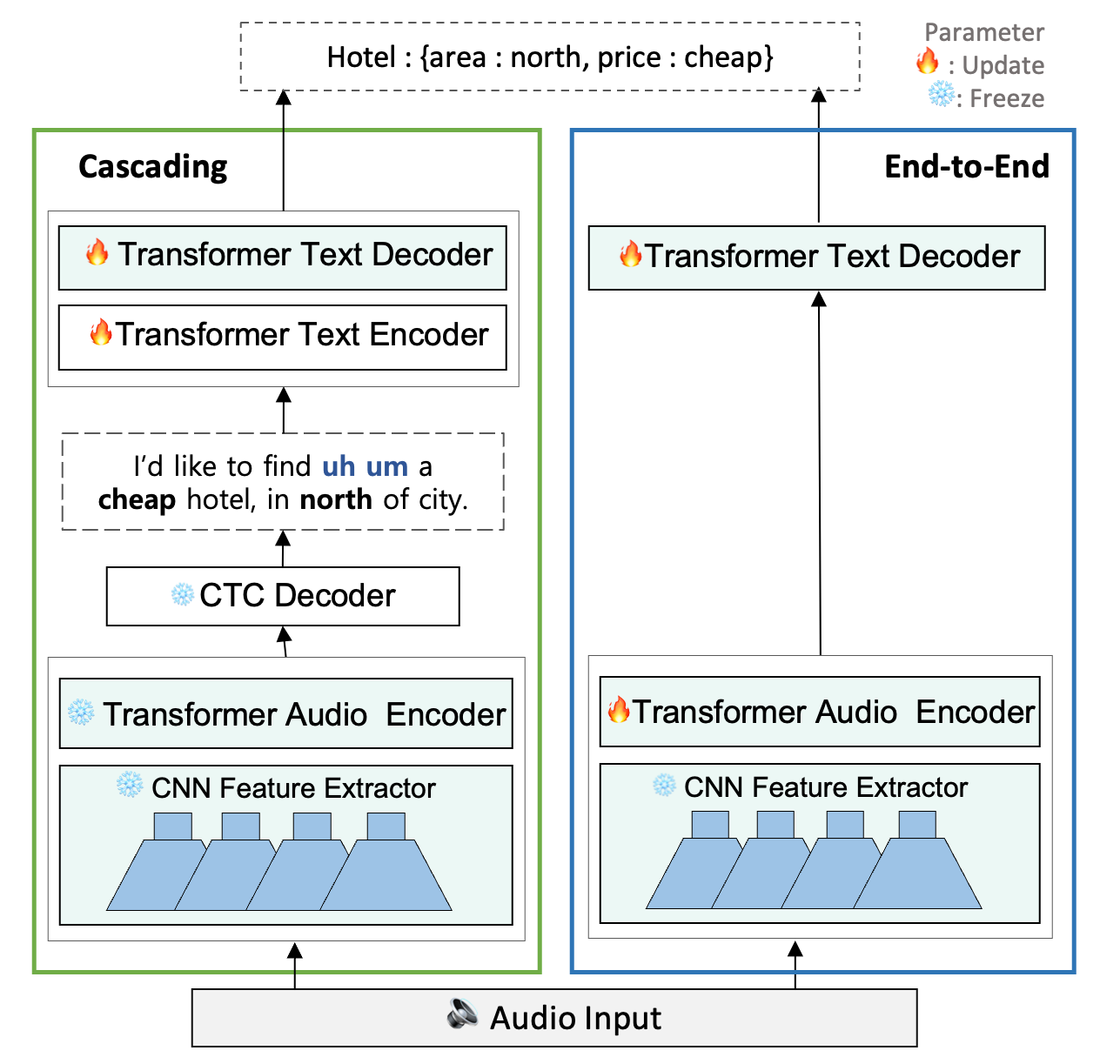}
  \caption{Schematic diagram of our cascading and E2E audio-based DST models. Cascading models utilize ASR transcripts, which may contain noisy text inputs. E2E bypasses ASR and directly utilize speech inputs.}
  \label{fig:model}
\end{figure}

\section{Methodology}

\begin{table*}[]
\centering

\scalebox{0.80}{ 
\begin{tabular}{p{52mm}|p{31mm}p{31mm}|p{8mm}p{18mm}}
\hline
Dialogue     & Gold & Prediction & F1 & \small{PhonemeF1} \\\hline
\small{{[}user{]} : I want the taxi to depart }
\newline from \textcolor{blue}{Stevanage}  to \textcolor{blue}{museum} for \textcolor{blue}{1} person. & \small{departure : \textcolor{red}{Stevanage} \newline destination : museum \newline bookperson : 1}    & \small{departure : \textcolor{red}{Stevanase} \newline destination : museum}   & 0.4  & 0.7989      \\\hline
\small{{[}user{]} : The restaurant I would reserve is \textcolor{blue}{Golden Wok}.}
 & \small{restaurant : \textcolor{red}{Golden} Wok}    & \small{restaurant : \textcolor{red}{Gordon} Wok}  & 0.0  & 0.9907      \\
\hline
\end{tabular}
}\caption{Comparison of F1 and PhonemeF1 scores with real examples.}
\label{tab:phonem_example}
\end{table*}


\subsection{Baselines}

\subsubsection{Cascading Model}

The cascading pipeline is comprised of two separate components: an ASR model that converts speech into text, and a transformer encoder-decoder model that functions as a belief state generator (Figure~\ref{fig:model}). Specifically, the ASR component is leveraged through W2V \cite{w2v} with Connectionist Temporal Classification (CTC) decoding. Meanwhile, a pre-trained BART \cite{bart} model that is trained with the denoising method is adopted as the belief state generator. After the ASR model transcribes input dialogue speech into text, the belief state generator takes the converted text-based input of dialogue turn $t$, which consists of previous system utterance ${s_{t-1}}$ and current user utterance ${u_t}$, and subsequently generates turn-level belief state $b_t$. We use negative log-likelihood as a loss function given $s_{t-1}$ and $u_t$, to train the belief state generator. Finally, the belief state $B_t$ at the turn $t$ is determined by accumulating $b_i$ from turns 1 to $t$. 

\subsubsection{End-to-End Model}

Unlike the cascading approach, E2E directly tracks the speech modality dialogue without ASR transcription (Figure~\ref{fig:model}). As such, this approach eliminates error propagation associated with ASR, and allows models to have access to intrinsic paralinguistic information (e.g., intonation) that is present in speech. Such additional information has the potential to facilitate user motive understanding, much like in actual human conversational interaction \cite{prosody}. To this end, we utilized the W2V model without CTC decoding as the speech encoder and incorporated a BART \cite{bart} model's decoder as the belief state generator. More precisely, unlike the cascading model, which uses the transcribed text as the input feature, E2E uses the embedded audio vector as the input feature for the belief state generator. Thus, the initial input of E2E at turn $t$ is of speech ${s_{t-1}}$ and $u_t$, and the model is trained to generate turn-level belief state $b_t$. Negative log-likelihood loss is used to simultaneously update the speech encoder and belief state generator. Similar to the cascading model, the belief state $B_t$ is derived by accumulating individual turn-level belief states
\subsubsection{Implementation Details}
For the W2V \cite{w2v} used in the cascading and E2E models, we used the pre-trained W2V-large checkpoint, which is identical to that in \cite{W2V-self} and is of 315M parameters. For BART \cite{bart}, we used a 140M parameter pre-trained model that has six bi-directional encoders and six auto-regressive decoders layers. During training, we used AdamW \cite{adamw} optimizer with a 5e-5 learning rate. Models are trained on A100 GPU with a batch size of 64 and a max epoch of 200.

\subsection{PhonemeF1 Metric}
\label{sec:metric}

We propose a new metric called \textbf{PhonemeF1} that offers a sophisticated evaluation of the predictions generated by audio-based DST models. While the F1 metric is widely used to assess text-based state tracking and SLU results \cite{trade, F1-cite2}, we present the PhonemeF1 metric to account for pronunciation similarity, which is crucial in analyzing correctness in spoken utterances within the audio modality. F1 calculates predictions in the following manner: when the model successfully predicts both the slot and its corresponding value, true positives (TP) are incremented. False-positives (FP) are counted when the predicted slot does not exist in gold answers, while false-negatives (FN) are added when the model skips and does not predict a specific slot\footnote{Evaluation script from https://github.com/jasonwu0731/trade-dst}. 

\begin{algorithm}[h]
\small
\caption{PhonemeF1 pseudo-algorithm}\label{alg:cap}
\textbf{Inputs: } Gold and predicted slot-value sequences $\mathcal{E}, \hat{\mathcal{E}}$ \\
$TP, FP, FN, common$ $\leftarrow$ 0 \\
$dist_{phone}$ $\leftarrow$ phonetic edit distance\\

\textbf{Output: } TP, FP, FN
\begin{algorithmic}[1]
\For{each $\hat{\epsilon} \in \hat{\mathcal{E}}$}
\If{$\hat{\epsilon} \in \mathcal{E}$}
\State $d = dist_{phone}(\hat{\epsilon}.value, {\epsilon}.value)$
\State $TP \mathrel{+}= (1 - d)$
\If {$\hat{\epsilon}.value == {\epsilon}.value$}
\State $common \mathrel{+}= 1$
\EndIf
\EndIf
\EndFor

\State $FP = len(\hat{\mathcal{E}})$ - $common$
\State $FN = len(\mathcal{E})$ - $common$
\end{algorithmic}
\end{algorithm}
On the other hand, F1 only counts exact word matches, so it fails to capture the nuances of phonetic variations seen in spoken utterances. In this context, the PhonemeF1 metric can be a better evaluation metric as it incorporates phonemic similarity between audio-based DST predictions and gold responses. In cases where slots are accurately predicted but values are partially incorrect, we first quantify pronunciation similarity between predicted and gold answers, and then add the degree of phonemic similarity as partial credit to the TPs present (Algorithm~\ref{alg:cap}). To calculate phonemic similarity, we first convert values into phonemic representations using the International Phonetic Alphabet (IPA), calculate phonetic Levenshtein edit distance\footnote{https://abydos.readthedocs.io/en/v0.4.1/\textunderscore modules/abydos/distance/\textunderscore phonetic \textunderscore edit \textunderscore distance.html}, and then subtract this distance from unity. Our PhonemeF1 metric not only provides nuanced scores, but also avoids excessive penalization of disparities that may result from ASR and audio representation in cascading and E2E models. For easier comprehension, an example is shown in Table~\ref{tab:phonem_example}.

\section{Experiments}

\subsection{Baselines, Evaluation Datasets, and Metrics} 


Towards audio-based DST, we implemented two different types of models: cascading and E2E. These models were exclusively trained on synthetic speech data. In order to evaluate their effectiveness in realistic environments, we tested our models on authentic human speech, and then compared the results with those conducted on synthetic test sets. To thoroughly gauge model performance, we utilized both F1 and our proposed PhonemeF1 metrics. In addition, to establish a benchmark and facilitate a comparison between audio-based DST models and existing text-based DST models, we implemented a BART-based textual model (Text-BART). It is worth noting that this text-based model is trained with error-free gold text \cite{MWOZ}, which is different from the ASR-transcribed texts that are used to train cascading models.


\begin{table}[t]
\centering

\scalebox{0.80}{
\begin{tabular}{cl|ll}
\hline
\multicolumn{1}{l}{TestSet} & Model    & F1     & PhonemeF1 \\\hline
\multirow{2}{*}{Human}        
& E2E   & 61.99 & 76.28  \\ 
& Cas.  & \textbf{62.07} & \textbf{76.79} 
\\\hline
\multirow{2}{*}{Synth}     
& E2E
& 67.59 \scriptsize{($\triangle$5.60)}
& 78.09 \scriptsize{($\triangle$1.81)}   \\  
& Cas. 
& \textbf{72.86} \scriptsize{($\triangle$10.79)}
& \textbf{82.72} \scriptsize{($\triangle$5.93)} \\\hline
Text & Text-BART \cite{bart} & 74.02 & 82.20\\\hline
\end{tabular}
}
\caption{Cas. and E2E comparisons on synthetic and human datasets, as well as text-based DST model. The $\triangle$ indicates the differences in scores between the two audio test datasets.}
\label{tab:main}
\end{table}

\subsection{Synthetic Audio Data For Authentic Human Speech}


In order to assess the viability of using synthetic data for training audio-based DST models, we tested our models on the human test set, and compared the results with the synthetic test split. Although Table~\ref{tab:main} indicates that training with synthetic datasets does not achieve perfect adaptation to authentic human speech as evidenced by discrepancies of $\triangle$5.60 and $\triangle$10.79 in F1, the incorporation of pronunciation similarity through the PhonemeF1 metric significantly reduces these score differences. Specifically, the cascading model demonstrated a diminished disparity of $\triangle$5.93, while the E2E model exhibited an even smaller difference of $\triangle$1.81. Moreover, when comparing our audio DST models with a text-based DST model that has been trained with gold texts, we find promising results; there is a marginal difference between the two modalities when considering phonetic similarity (E2E: $\triangle$5.92, Cascading: $\triangle$5.41 on the human test set). These results in Table~\ref{tab:main} suggest that despite not achieving an exact correspondence between the predicted and gold values, the utilization of a synthetic training dataset allows for more effective discernment and tracking of the intended words, generating responses that bear notable resemblances in pronunciation to the ground truth.



\subsection{Cascading Versus E2E}
The cascading model demonstrates superior performance compared to the E2E model when evaluated on the human test dataset (Table~\ref{tab:main}). This can be attributed to speech to text conversion via ASR, which mitigates miscellaneous information such as background ambient noise. However, the difference in F1 between the two approaches is not substantial (E2E: 61.99, Cascading: 62.07). Furthermore, the score differences between the human and synthetic test set is smaller for the E2E ($\triangle$5.60) compared to the cascading model ($\triangle$10.79). This implies that despite requiring less parameters compared to its cascading counterpart, E2E is also a viable option when training with synthetic data.

\subsection{Multi-Speaker Versus Single-Speaker}

The SynthWOZ dataset is curated with multiple speakers to enhance the robustness and generalizability to authentic human audio. To investigate the effectiveness of leveraging multiple speakers, we created a separate single-speaker version of SynthWOZ, and trained the E2E model on this dataset. As evident from the results in Table~\ref{tab:single_speaker}, it is clear that the model trained on the multi-speaker SynthWOZ dataset generally outperformed its single-speaker counterpart across both the F1 and PhonemeF1 metrics. For the human test set, F1 scores increased from 51.88 to 61.99, and PhonemeF1 increased from 67.44 to 76.28 when using the multi-speaker setting. Furthermore, the multi-speaker setting significantly decreases the difference between the results of the synthetic and human test set from $\triangle$15.77 to $\triangle$5.60 with F1. Consequently, the widely recognized understanding that diverse data plays a pivotal role in enhancing model robustness remains pertinent in our specific context; incorporation of diverse multi-speaker datasets is a critical factor in facilitating models' ability to generalize to actual real-life circumstances.


\subsection{Multi-Turn Versus Single-Turn}

The task of audio-based DST involves multi-turn conversations between a system and a user, wherein the system's utterances play a crucial role in providing meaningful context for comprehending the conversation's progression. In order to effectively harness this system context, we additionally incorporated synthetic audio of the system's utterances into the SynthWOZ dataset. To evaluate the impact of utilizing system audio when training audio-based DST models, we conducted experiments involving two scenarios: one involving both user and system audio (S + U), and the other involving only user audio (U). The summarized results in Table ~\ref{tab:with_sys} demonstrate that the inclusion of system context yields significant enhancements compared to models solely relying on user audio; F1 scores increase from 56.84 to 61.99 for the E2E model, and from 52.16 to 62.07 in the cascading model. Similar trends can be observed with the PhonemeF1 metric. Furthermore, by using the multi-turn dataset instead of the single-turn version, the differences between the human and synthetic test datasets decrease for both the cascading and E2E models. 



\begin{table}[]
\centering

\scalebox{0.85}{
\begin{tabular}{ll|ll}

\hline
TestSet & Speaker & F1      & PhonemeF1        \\ \hline
\multirow{2}{*}{Human}  
& Single      & 51.88 & 67.44 \\
& Multi    & \textbf{61.99} & \textbf{76.28} \\ \hline
\multirow{2}{*}{Synth}
&Single
&\textbf{67.65} \scriptsize{($\triangle$15.77)}
&76.00 \scriptsize{($\triangle$8.56)}     
\\ 
&Multi
&67.59 \scriptsize{($\triangle$5.60)} 
&\textbf{78.09} \scriptsize{($\triangle$1.81)}        
\\\hline
\end{tabular}
}
\caption{Single- and multi-speaker datasets generalizability.}
\label{tab:single_speaker}
\end{table}

\begin{table}[]
\centering

\scalebox{0.74}{
\begin{tabular}{cl|l|ll}
\hline
\multicolumn{1}{c|}{Test} & \multicolumn{1}{c|}{Model}  & \multicolumn{1}{c|}{Input}& \multicolumn{1}{l}{F1} & \multicolumn{1}{c}{PhonemeF1} \\ \hline
\multicolumn{1}{c|}{\multirow{4}{*}{Hum.}} & \multirow{2}{*}{\small{E2E}} 
& U                  
& 56.84                
& 69.51

\\

\multicolumn{1}{c|}{} &                      

& \footnotesize{S+U}    
& \textbf{61.99}
& \textbf{76.28}
\\\cline{2-5} 
\multicolumn{1}{c|}{}                     
& \multirow{2}{*}{\small{Cas.} }
& U
& 52.16
& 68.33                    
\\
\multicolumn{1}{c|}{} &                      
& \footnotesize{S+U}    
& \textbf{62.07}
& \textbf{76.79}
\\\hline

\multicolumn{1}{c|}{\multirow{4}{*}{ Syn.}} & \multirow{2}{*}{\small{E2E}}   
& U                     
& 64.60 \scriptsize{($\triangle$7.76)}                 
& 74.89 \scriptsize{($\triangle$5.38)}                     \\ 
\multicolumn{1}{c|}{}                     &                  
& S+U     
& \textbf{67.59} \scriptsize{($\triangle$5.60)}                 
& \textbf{78.09} \scriptsize{($\triangle$1.81)}                       \\
\cline{2-5} 
\multicolumn{1}{c|}{}                     & \multirow{2}{*}{\small{Cas.}} 
&U               
& 69.84 \scriptsize{($\triangle$17.68)}                 
& 80.04 \scriptsize{($\triangle$11.71)}\\
\multicolumn{1}{c|}{}                     &                      
& S+U     
& \textbf{72.86} \scriptsize{($\triangle$10.79)}                 
& \textbf{82.72} \scriptsize{($\triangle$5.93)}   \\                  
\hline
\end{tabular}
}
\caption{Effect of contextual information.}
\label{tab:with_sys}
\end{table}



\section{Analysis}

\subsection{Error Analysis}

\begin{figure}[!h]
\begin{minipage}[t]{0.43\linewidth}
    \centering
    \includegraphics[width=1\textwidth]{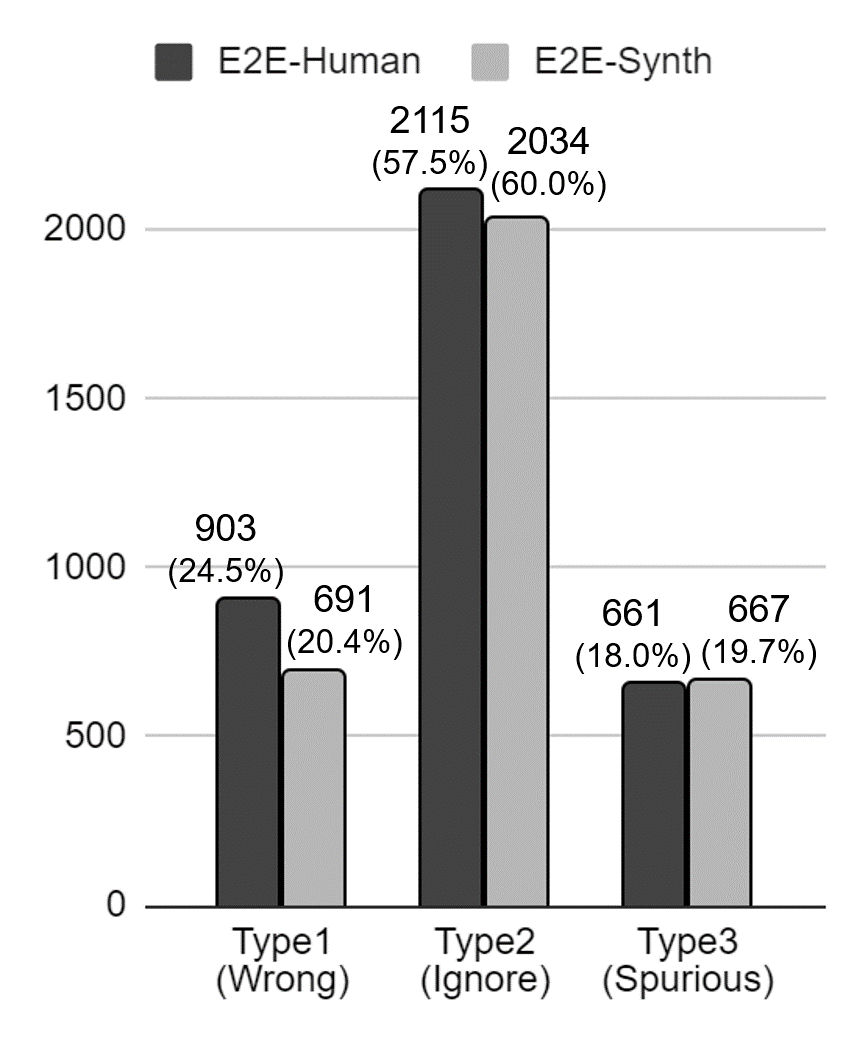}
    \caption{E2E. error types.}
    \label{fig:error_e2e}
\end{minipage}
\hspace{0.01cm}
\begin{minipage}[t]{0.43\linewidth} 
    \centering
    \includegraphics[width=1\textwidth]{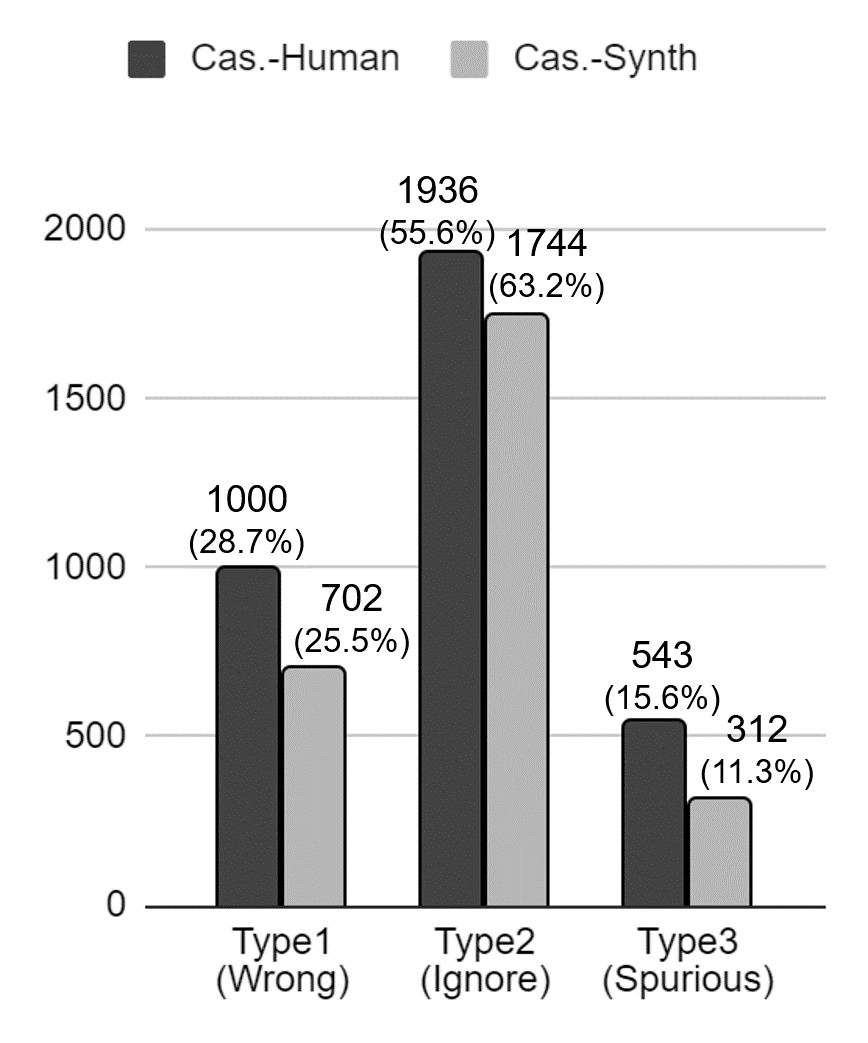}
    \caption{Cas. error types.}
    \label{fig:error_cas}
\end{minipage}        
\end{figure}



We conduct analysis of the predictions made by models trained on the synthetic dataset, wherein we classify the errors into three distinct categories. The classification results for the E2E and cascading models are presented in Figure~\ref{fig:error_e2e} and Figure~\ref{fig:error_cas}, respectively. The y-axis indicates the number of incorrect predictions, while the percentages in parentheses indicate the ratio of each error type per model. Specifically, Type 1 errors correspond to cases in which the model correctly identifies the slot, but fails to predict the exact values. Type 2 errors occur when the model neglects to predict a mentioned slot altogether. Type 3 errors encompass situations where the model generates spurious values that are not mentioned within the dialogue.


These findings yield valuable insights. First, the Type 2 errors emerge as the primary source of inaccuracies for both the E2E and cascading models. This implies that in order to ensure the recognition of informative user utterances, a more diverse range of scenarios is required within the synthetic dataset. Second, the most notable discrepancy in error types between the human and the synthetic test set manifests as Type 1 errors. Further analysis shows that among the Type 1 errors, about 9\% had minor phoneme edit distance when compared with gold answers\footnote{Less than 0.02 for 8.98\% in the E2E and 9.90\% in the cascading model.}. This means that there are only subtle pronunciation variations between predicted and ground truth values. While not the predominant factor, future research could benefit from improving the diversity of the synthetic dataset to include a broader range of pronunciations. 




\subsection{Accuracy Per Answer Type}


\begin{table}[]
\centering
\scalebox{0.77}{
\small{
\begin{tabular}{ll|p{10mm}p{10mm}p{10mm}p{10mm}}
\hline
\multicolumn{1}{l|}{Test} & Model& \multicolumn{1}{p{10mm}}{Name \newline \footnotesize{(40.8\%)}} & \multicolumn{1}{p{10mm}}{Cat. \footnotesize{(33.25\%)}} & \multicolumn{1}{p{10mm}}{Number \footnotesize{(16.1\%)}} & \multicolumn{1}{p{10mm}}{Time \footnotesize{(9.8\%)}} \\ \hline
\multicolumn{1}{l|}{\multirow{2}{*}{Hum.}}
& E2E 
& 34.61 & \textbf{59.56} & \textbf{40.54} & 42.22 \\
\multicolumn{1}{l|}{}                              
& Cas.
& \textbf{42.04} & 50.39 & 39.43  & \textbf{43.52}  \\\hline

\multicolumn{1}{l|}{\multirow{4}{*}{Syn.}} 
& \footnotesize{E2E}
& 40.91 \scriptsize{($\triangle$6.30)}                                      
& \textbf{61.31} \scriptsize{($\triangle$1.75)}                                       
& 46.13 \scriptsize{($\triangle$5.59)}                                       
& 50.40 \scriptsize{($\triangle$8.18)} 
\\

\multicolumn{1}{l|}{}                              
& Cas.
&\textbf{54.21} \scriptsize{($\triangle$12.17)}                                      
&55.37 \scriptsize{($\triangle$4.98)}                                       
& \textbf{51.88} \scriptsize{($\triangle$12.45)}                                        
& \textbf{54.22} \scriptsize{($\triangle$10.7)}   
\\\hline

\end{tabular}
}
}\caption{Accuracies for different answer types.}
\label{tab:cat_error}
\end{table}

We conduct additional examination to assess the accuracy of different answer types on both human and synthetic datasets. The predicted values for each slot were systematically classified into four distinct categories: Name, Categorical, Number, and Time. The distribution of these categories is presented in Table~\ref{tab:cat_error}. Notably, the cascading model exhibited better performance in the Name category compared to the E2E model, achieving accuracies of 42.04 and 34.61, respectively for the human test set. This discrepancy emphasizes the effectiveness of integrating ASR transcription, which can help accurate tracking of relatively complex proper nouns. In contrast, the E2E model demonstrates superior results in the Categorical answer types (59.56) compared to the cascading model (50.39), and there is minimal disparity between the results of the human and synthetic test sets ($\triangle$1.75). These findings suggest that direct audio inputs facilitate a more straightforward optimization process for categorical values.


\section{Conclusion}
In this paper, we expand the text-based DST task into the audio domain. Specifically, we curate the SynthWOZ dataset, to explore the viability of employing synthetic audio data as a realistic alternative to authentic human recordings. Through experimentation and analyses, we compare our SynthWOZ dataset with a separately curated human audio dataset, while also introducing cascading and E2E baselines. Moreover, we introduce a novel PhonemeF1 metric that considers pronunciation similarity, which offers a more reasonable evaluation for audio-based interactions. We anticipate that our investigation will prove invaluable to the audio DST research community.


\section{Acknowledgements}
\begin{small}
Supported by Institute of Information \& communications Technology Planning \& Evaluation grant funded by the Korea government (No.2019-0-01906, Artificial Intelligence Graduate School Program (POSTECH) and No.2021-0-00575, Development of Voicepishing Prevention Technology Based on Speech and Text Deep Learning).
\end{small}
\bibliographystyle{IEEEbib}
\bibliography{strings,refs}

\end{document}